# Management and Detection System for Medical Surgical Equipment

Alexandra Hadar
*Faculty of Industrial Engineering and Technology Management*
*Holon Institute of Technology - HIT*
Holon, Israel
msalexandrahadar@gmail.com

Natan Levy
*School of Computer Science & Engineering*
*The Hebrew University of Jerusalem*
Jerusalem, Israel
Natan.levy1@mail.huji.ac.il

Michael Winokur
*Faculty of Industrial Engineering and Technology Management*
*Holon Institute of Technology - HIT*
Holon, Israel
Michaelw@hit.ac.il

*Abstract*—Retained surgical bodies (RSB) are any foreign bodies left inside the patient after a medical procedure. RSB is often caused by human mistakes or miscommunication between medical staff during the procedure. Infection, medical complications, and even death are possible consequences of RSB, and it is a significant risk for patients, hospitals, and surgical staff. In this paper, we describe the engineering process we have done to explore the design space, define a feasible solution, simulate, verify, and validate a state-of-the-art Cyber-Physical System that can significantly decrease the incidence of RSB and thus increase patients' survivability rate. This system might save patients' suffering and lives and reduce medical staff negligence lawsuits while improving the hospital's reputation. The paper illustrates each step of the process with examples and describes the chosen solution in detail.

I. INTRODUCTION

Today the complexity of the modern healthcare system is growing alongside the population growth in such a way that requires the provision of treatments to more patients with less staff. The workload on the medical staff in surgeries is increasing. Medical equipment is reconciled manually in most hospitals during surgeries. Retained Surgical Bodies (RSB) can cause infections, severe medical complications, and even death [1]. This event is also described in the literature as Forgotten Foreign Bodies (FFB) [2]. The incidence of this condition is between 0.3 and 1.0 per 1,000 abdominal operations. In the United States, for example, there are about 1,500 RSB cases per year [1]. According to a recent national survey in the United States, retained sharp instruments (needle, blade, guidewire, metal fragment) are more prevalent than reported in the current literature [3]. Standardizing reports and implementing new technologies is the most effective way to improve the management and prevention of these events [4].

Internet of Things (IoT) technology can provide solutions that allow medical staff to track medical equipment and prevent RSB. In this work, we describe the design of a Cyber-Physical System, centered around Radio Frequency Identification (RFID) based IoT system that will highly decrease the incidence of RSB. This technology can reduce the need for unnecessary secondary surgical procedures and increase patients' survivability rate.

There is evidence that technologies like RFID may help to reduce the occurrence of RSB [5]. The use of RFID sponge detection technology reduced the percentage of procedures in which a search for a sponge was performed, the number of unreconciled sponge counts, the amount of time spent searching for sponges and obtaining radiographs, and costs [6].

There is medical equipment marked with RFID and barcode, as well scanning and tracking systems with RFID, like Xarefy [7] and ORLocate [8], and barcode like Stemato [9].

In those systems, there is no automatic detection of the medical equipment at the room entrance. There is a need to bring the detector near to the medical equipment. Moreover, in the case of a barcode, there is a need for a line of sight. There is no registration and automatic alerts for all cases where medical equipment leaves or enters the operating room during surgery.

Except for ORLocate, the above systems do not have a dedicated component for locating equipment in the patient's cavity.

This paper manuscript focuses on the design challenges we faced to achieve an optimal feasible solution. Further development is needed to implement a full prototype to perform controlled experiments to rate the success of the proposed design in the field. Section II presents an overarching view of the process, Section III describes in detail the design space exploration performed to reach the desired solution, and section IV describes the selected architecture including steps taken during conceptual design to augment the system's robustness. Section V describes the solution validation. The paper's concluding summary and directions for future work are presented in Section VI.

II. WORK PROCESS

The authors started by looking at existing technologies. Next, the stakeholder disclosure process was done, and a collection of their needs and requirements, including detailed interviews with clinical staff, reviews of news media, and professional literature.

In all the interviews, the interviewees expressed their concerns about the current situation. The majority of the interviewees believe that a system is needed to solve those concerns. The main topics that were raised:

• The process of counting medical equipment in most hospitals is currently done manually - exposing the process to human error.

• It is necessary to reduce the load on the medical staff.

• The proposed solution must include a way of human intervention.

• Concerns have been raised about the level of disruption the system may cause to the medical staff (physical, visual, vocal), concerns about unnecessary system movement that could interfere with the proper course of surgery and damage the sterile field.

• Medical staff may have difficulty using a cumbersome solution.

• To avoid unnecessary interference, the solution should be operated "hands-free" via voice commands and gestures, rather than only by buttons.

In conclusion, there is a need to facilitate the work processes of the medical staff and help prevent human error. There is a need for a solution with minimal intervention in the existing process, without unnecessary movements that can impair the surgical process and the sterile field.

Next, we sorted the needs according to the KANO method [10] and rated them according to the Nominal Group Technique (NGT) method [11]. As a result, the six top fundamental requirements are: (1) documenting and reporting the presence, (2) counting, (3) monitoring and identifying the medical equipment, (4) identifying the room where the medical equipment is present, (5) monitoring the medical equipment in the operating room, and (6) locating medical equipment in the operating space.

The following stages included the definition of system use cases (Fig. 1 presents an example of a central use case of locating medical equipment within the patient's cavity once medical staff announce patient closing), system requirements, engineering characteristics of the possible solutions, and building a system model, including its dynamics (Fig. 2).

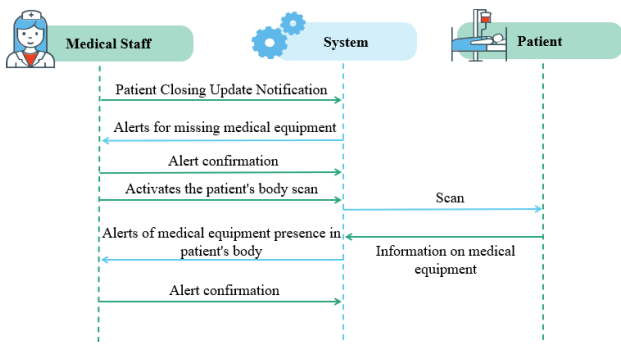

Fig. 1. Use Case - Locating medical equipment in the patient's cavity

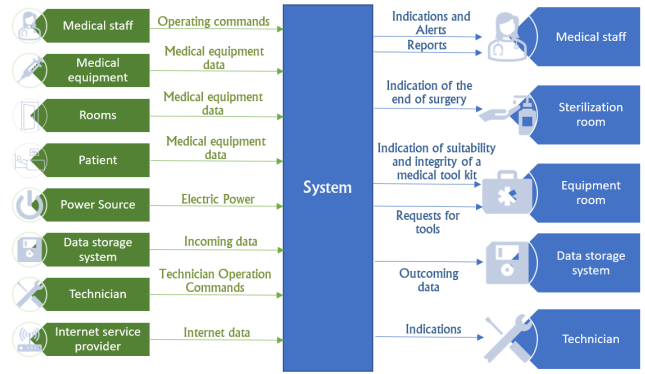

Fig. 2. System model with its dynamics

Seven different solution concepts were proposed, to systematically select the most appropriate solution. We performed a comprehensive failure modes analysis for the selected solution and suggested new features to improve system robustness.

Next, we defined the requirements for the subsystem components and established a detailed system Verification and Validation (V&V) process, including a detailed simulation of the system dynamics with MathWorks Simulink [12] to check the design validity and robustness.

III. DESIGN SPACE EXPLORATION AND SYSTEM SELECTION

The design space exploration performed started with the established principles of concept definition of possible solutions followed by the generation of design variants and their evaluation to reach a selected "best alternative" [13]. This exploration included a combination of different technologies to monitor the medical equipment and locate it in the operating space via, for example, RFID, Bluetooth, Ultra-Sound, and Cameras. Some solutions included robots to handle the equipment while other solutions included a mobile cart with portable detectors for this task.

*A. Solution Variants*

A morphological matrix as presented in Table I was applied to generate five solutions variants and perform a risk assessment for components in each solution. We proposed two more combinations of solutions using mix & match principles [13]. Concepts short description:

*1) Dr. Tool:* RFID-based equipment tracking system. Built-in RFID sensors into all medical equipment. Scattered sensors in the rooms' entrances (Room Sensors) enable monitoring of the location of the equipment in the room area. Mobile Tool Cart (MTC) RFID-based medical tool cart with monitor, computer and communication system, RFID tool tray, and RFID trash bin. Medical Equipment Detector (MED) RFID detector for manual scanning or mounted to surgical lights pivoting arm to locate misplaced equipment. Cloud-based Central Management System (CMS) for database management of the medical equipment, receiving an indication from the various sensors, and generating reports.

TABLE I. THE MORPHOLOGICAL MATRIX

| Main Functions | Concepts | | | | | | |
|---|---|---|---|---|---|---|---|
| | *Dr. Tool* | *Blue Tool* | *Ultra Tool* | *Robi Tool* | *BB Tool* | *Dr. Robi Tool Mix & Match* | *Dr. RoBBi Tool Mix & Match* |
| Monitoring of medical equipment | RFID-based system: RFID sensors built into medical equipment, RFID Room Sensors, MTC, and MED | Integrated Bluetooth and RFID system: Bluetooth sensors built into medical tools, RFID sensors built into consumable equipment, Bluetooth and RFID-based medical equipment vending machine, RFID Room Sensors, MTC, and MED | RFID-based system: RFID sensors built into medical equipment, RFID Room Sensors, MTC, and MED | RFID-based system with a robot: RFID sensors built into medical equipment, RFID Room Sensors, Robot with built-in RFID readers | Integrated system based on RFID and cameras: cameras with lighting and medical equipment detection algorithm, RFID sensors built into medical equipment, RFID Rooms Sensors, MTC, and MED with cameras | RFID-based integrated system with a robot: A system of RFID sensors built into medical equipment, RFID Rooms Sensors, Robot with built-in RFID readers | RFID-based integrated system with a robot and cameras: Camera system with lighting and medical equipment detection algorithm, A system of RFID sensors built into medical equipment, RFID Rooms Sensors, Robot with built-in RFID readers |
| Identifying equipment in the patient cavity | MED | Bluetooth for the medical tools, RFID sensors for the consumable medical equipment with MED | Dedicated ultrasound monitoring system in the patient's bed | Surgery bed with RFID readers | IR cameras system with medical equipment detection algorithm, thermography | MED | MED |
| Providing alerts and indications | Alerts and indication system with monitor and speaker on the MTC's local computer, sounds and indication lights on MED | Alerts and indication system on smartphone connected to the MTC's monitor | Alerts and indication system with monitor and speaker on the MTC's local computer, Ultrasound system with sound and indication lights | Alerts and indication system in the robot's tablet, sounds and indication lights on the surgery bed | Alerts and indication system on the MTC's tablet | Alerts and indication system in the robot's tablet, sounds and indication lights on MED | Alerts and indication system in the robot's tablet, sounds and indication lights on MED |
| Communicating with medical staff | Communication system[1] Installed on MTC's local computer | Communication system[1] Installed on smartphone connected to the MTC's monitor | Communication system[1] Installed on MTC's local computer | Communication system[1] Installed on robot | Communication system[1] Installed on MTC's tablet | Communication system[1] Installed on robot | Communication system[1] Installed on robot |
| Task management and generating reports | CMS with SQL DB + Python | CMS with SQL DB + Matlab | CMS with Cassandra DB + Python | CMS with Oracle DB + Python | CMS with Oracle DB + Matlab | CMS with SQL DB + Python | CMS with Oracle DB + Matlab |
| Saving data and history | Cloud | Cloud | External backup drive | Internal backup drive | Company servers | Cloud | Cloud |

[1.] Communication system with command decoding - voice decoding software with microphone, speakers, modem, camera, and hand gesture decoding software

*2) Blue Tool:* Bluetooth and RFID-based integrated system. Bluetooth sensors built into tools and RFID built into consumable medical equipment, RFID Room Sensors. Medical equipment vending machine based on Bluetooth and RFID, issuing medical equipment according to a command from MTC or CMS, smartphone-based MTC, and MED for the consumable misplaced medical equipment, Cloud-based CMS.

*3) Ultra Tool:* Ultrasound and RFID-based system with built RFID sensors inside all the medical equipment. RFID Room Sensors, MTC, and MED for locating medical equipment in the operating room space, a dedicated ultrasound system for identifying equipment in the patient cavity, and CMS with an external backup drive.

*4) Robi Tool:* RFID-based system equipment tracking system with a robot. RFID sensors built into all medical equipment, RFID Room Sensors, a robot with a built-in tablet, communication system, RFID tool tray, RFID trash bin, and scattered sensors in the robot frame for locating equipment. The robot can receive tools from the equipment room, move them to the operating room and deliver them to medical staff. A dedicated surgical bed equipped with RFID readers for identifying equipment in the patient cavity and CMS with an internal backup drive.

*5) BB Tool:* Integrated RFID, cameras with a lighting system, and Infra-Red (IR) cameras. Cameras and RFID Room Sensors enable monitoring of the location of tools in the room. Cameras enable proactive detection of missing tools in the room space. Tablet-based MTC and MED, in addition to the MED containing cameras for detection in a wider visible range in the operating room space. IR cameras for identifying equipment in the patient cavity using thermography, and CMS with storage on company servers.

*6) Dr. Robi Tool (Mix&Match):* Combination of Dr. Tool & Robi Tool. RFID Room Sensors, Robi Tool Robot, MED, and Cloud-based CMS.

*7) Dr. RoBBi Tool (Mix&Match):* Combination of Dr. Tool & Robi Tool & BB Tool. Cameras and RFID Room Sensors, Robi Tool Robot, MED, and Cloud-based CMS.

### B. Evaluating Main Solution Variants

To find the best solution, we used the PUGH method [14]. In this section, we describe the process to find the three most suitable candidates using Engineering and Qualitative Characteristics and refined the choice to three possible solutions. The preferred solution is then described in Section IV below.

*1) Engineering characteristics*

The essential mechanism for solution selection is the engineering characteristics that should be common to all of them [13]. Table II presents the technical engineering characteristics that were selected for solutions comparison.

We calculated the relative importance of each engineering characteristic using the "Voice of the Customer" aspects of the Quality Function Deployment (QFD) method [15] by the quantitative analysis of the correlation between characteristics and stakeholders' needs. The top five highest scoring characteristics were used for an initial evaluation of all the solutions (Availability, Detection Range, Reliability – MTBF, Charging Time, Screen Size).

*2) Qualitative characteristics*

We apply a slight variation of classical QFD for solution ranking for the sake of clarity, using a two-dimensional diagram where the engineering characteristics are used for technical ranking on the Y axis, while on the X axis we display the ranking of "qualitative" characteristics as applied in [16]. It is shown in Fig 3. The qualitative characteristics for solutions comparison that were selected are Time to Market, Life Cycle Cost, End User, Dependence on Suppliers, Reliability, Availability, Maintainability, and Safety (RAMS).

TABLE II. ENGINEERING CHARACTERISTICS TABLE

| Characteristic Name | Characteristic Data | | |
|---|---|---|---|
| | *Type* | *Range* | *Target* |
| Height | Quantitative | (0.85 to 1.20) m | (0.85 to 1.15) m adjustable |
| Diameter | Quantitative | (0.60 to 0.75) m | (0.67 to 0.71) m adjustable |
| Weight Carrying | Quantitative | (15 to 18) kg | 18 kg |
| Availability | Quantitative | (98 to 99) % | 98 % |
| Maximum Humidity | Quantitative | (80 to 90) % | 90 % |
| Minimum Humidity | Quantitative | (0 to 10) % | 5 % |
| Maximum Temperature | Quantitative | (40 to 50) °C | 40 °C |
| Minimum Temperature | Quantitative | (-5 to 1) °C | 1 °C |
| Noise Level | Quantitative | (65 to 85) dB | 80 dB |
| Charging Time | Quantitative | (9000 to 18000) s | 9000 s |
| Screen Size (diagonal) | Quantitative | (0.5461 to 0.6858) m | 0.6858 m |
| Detection Range | Quantitative | (0.2 to 1) m | 0.9 m |
| Reliability - Mean time between failures (MTBF) | Quantitative | (4320000 to 5400000) s | 5184000 s |

*3) Top Three concepts*

After performing an initial evaluation table (screening) we found that the attempt to use robots to handle the medical equipment in cooperation with surgical personnel did not pass the initial evaluation stage in the PUGH method. Robots are a more complex alternative compared to the medical cart with scattered sensors and detectors with RFID and are still considered a less reliable option by skilled personnel.

The top three alternatives are Dr. Tool, Ultra Tool, and Blue Tool.

*4) The best concept*

After conducting a comprehensive scoring evaluation for the top three variants in a relation to all the technical engineering characteristics, and the summary of the weighted scores for each of the alternatives, we found that Dr. Tool's alternative is the top concept. We performed a qualitative characteristics rating for the top three solutions. The results of the technical characteristics scoring and the qualitative characteristics comparison appear in Fig. 3.

From the comparison of solution alternatives, using both technical and qualitative ratings, Dr. Tool is the best concept.

## IV. SELECTED CONCEPT

### A. Solution Robustness

Before we can target a final state-of-the-art design concept, we applied techniques to check and enhance the robustness of the solution since this is a key to its success. For example, see [13, 17]. We used NASA's Risk Management Matrix [18] to identify weaknesses and potential failures in the selected option in order to reduce risks. Risk reduction is achieved thru redundant sensors and additional batteries, reinforcements to the tray, lowering the center of mass, and improving cyber security. Consequently, the robust Dr. Tool concept presented in Section B below got the highest score with PUGH reevaluation.

Reevaluation of the results of the technical characteristics scoring and the qualitative characteristics comparison with the robust solution appears in Fig. 4.

### B. Final Concept

The design of the final concept is presented schematically in Fig. 5. The architecture includes the following components:

1) *Built-in RFID sensors in medical equipment.*
2) *Room Sensors:*
   Rooms are equipped with an RFID-based tracking system to monitor medical equipment location within the equipment room, Sterile Processing Department (SPD), and operating room.

3) *Mobile Tool Cart (MTC):*
   The central part of the innovative design. Portable RFID-based medical cart with a monitor and computer. The MTC can communicate with medical staff using voice commands, gestures, and a touch screen. It is equipped with an RFID tool tray and an RFID trash bin for tracking medical equipment.

4) *Medical Equipment Detector (MED):*
   RFID detector to locate misplaced medical equipment in the patient cavity or operating room. The detector can be used for manual scanning or mounted to surgical lights pivoting arm.

5) *Central Management System (CMS):*
   Central Cyber component of the system. Cloud-based service system responsible for communicating with all room sensors and MTC, creating alerts to the medical staff, managing the medical equipment database, and generating reports.

This system can identify each medical equipment item unambiguously, including needles. The MED can be installed above the operating space. The MTC supports voice commands and hand gestures in addition to a touch screen. The strengths of the system are the ability to locate medical equipment automatically at the entrance to the rooms, and it will automatically notify the presence of the medical equipment during the operation. When the patient's cavity is about to close the system will notify the medical staff for RSB.

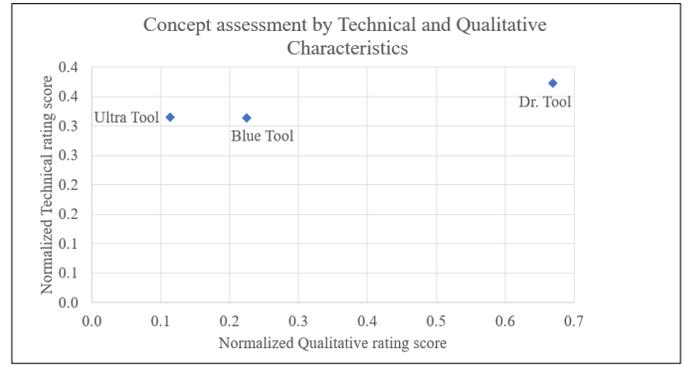

Fig. 3. Comparison of the technical and qualitative ratings of the solution concepts

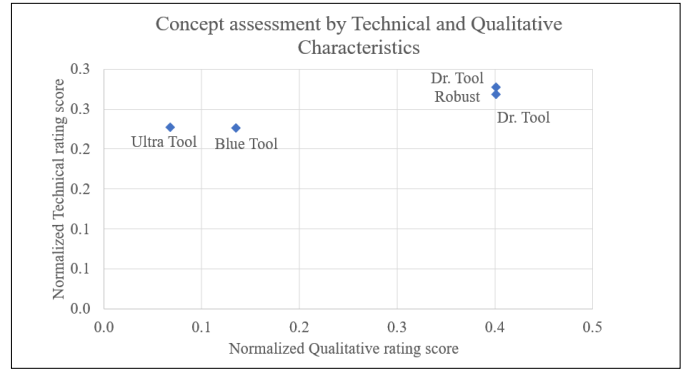

Fig. 4. Comparison of the technical and economic ratings of the solution concepts with robust Dr. Tool

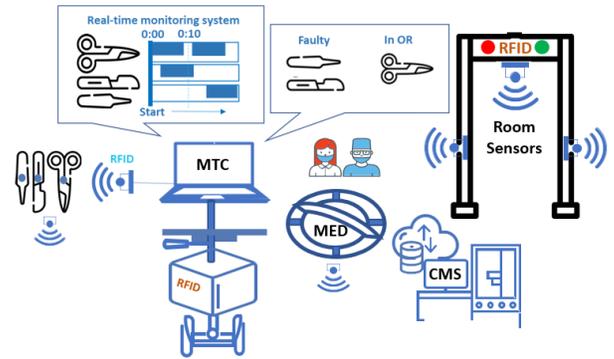

Fig. 5. System Design

## V. SYSTEM SIMULATION AND VERIFICATION

An early-stage Verification and Validation (V&V) process was applied to check the system design. We developed a model of the system and its dynamics, and a state-flow simulation was performed on the main processes using Simulink [3]. This includes monitoring the medical equipment in the operating room during surgery, including patient cavity, medical equipment room locator, medical staff reporting in case of a damaged item, and generating reports. Part of the simulation for monitoring medical equipment during surgery appears in Fig. 6.

During simulation, new inputs were added, such as acknowledgment by the SPD that they are ready to receive contaminated instruments at the end of the operation.

The simulation highlighted the relevance of communication between the CMS and MTC in reporting the status of the medical equipment presence in the Operation Room (OR).

In case new medical equipment is brought into the OR without placing it on the MTC, the CMS will notify the MTC about the presence of new equipment, and the MTC will add it to the monitoring list and alert the medical staff.

All medical equipment is being monitored constantly during operation, even if the medical staff accidentally forgot a tool in his pocket and left the OR, CMS will notify the MTC, and MTC will remove this tool from the monitoring checklist for this surgery and will inform the medical staff. After performing the simulation multiple times, with different sequences and inputs, we got a high level of trust that the design meets the stakeholders' requirements.

## VI. Conclusions

Technology is evolving and advanced IoT systems today allow us to improve existing processes and increase automation. In our work, we have taken a step towards automation in the field of monitoring medical equipment - in a hospital that will save lives. However, the most suitable design we have chosen is not completely autonomous, as it still requires a final scanning performed by trained medical personnel to have full confidence of non-residual RSB.

We argue that at this stage the best way to enter the market is with a less complex and more reliable system that the medical staff can incorporate into the standard procedure in the hospital. Future development envisages the construction of a working prototype and the performance of controlled experiments to gather data supporting its validity. Future versions of the system should incorporate robotic handling of the surgical equipment and AI. In a collaborative environment with medical personnel, this will further streamline and fully automate the process of monitoring the medical equipment to minimize the risk of RSB.

## Acknowledgment

We would like to thank Mr. Yossi Hen for his originality, intense work investment, and partnership in the design phase of the project.

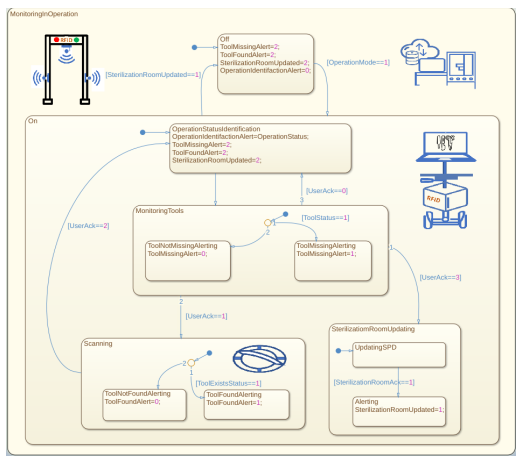

Fig. 6. Part of State-Flow simulation in Simulink